\documentclass[10pt,a4paper]{article} 

\usepackage[margin=1in]{geometry}
\usepackage{slashbox}

{}
{}
{}
\usepackage{url}
\usepackage{enumerate}
\usepackage{graphicx}
\usepackage{braket}
\usepackage{todonotes}
\usepackage{amsmath}
\usepackage{hyperref}
\usepackage{mathrsfs}
\usepackage{amsfonts}
\usepackage{mathtools}
\usepackage{comment}
\usepackage{longtable}
\usepackage{multicol}
\usepackage{multirow}

\usepackage{wrapfig}
\usepackage{lscape}
\usepackage{rotating}
\usepackage{makecell}
\usepackage{authblk}
\usepackage{subcaption}

\title{Efficient Decoding of Surface Code Syndromes for\\ Error Correction in Quantum Computing}

\author[1,*]{Debasmita Bhoumik}

\author[2]{Pinaki Sen}

\author[1]{Ritajit Majumdar}

\author[1,+]{Susmita Sur-Kolay}
\author[3]{ Latesh Kumar K J}

\author [3]{ Sundaraja
Sitharama Iyengar}

\affil[1]{Advanced Computing \& Microelectronics Unit, Indian Statistical Institute, Kolkata, India}
\affil[2]{Department of Electrical Engineering, National Institute of Technology, Agartala, India}

\affil[3]{KFSCIS, Florida International University, Miami, Florida,USA}

\affil[*]{debasmita.ria21@gmail.com}
\affil[+]{ssk@isical.ac.in}

\date{}
\begin{document}
\maketitle

\begin{abstract}
Errors in surface code have typically been decoded by Minimum Weight Perfect Matching (MWPM) based method. Recently, neural-network-based Machine Learning (ML) techniques have been employed for this purpose. Here we propose a two-level (low and high) ML-based decoding scheme, where the first level corrects errors on physical qubits and the second one corrects any existing logical errors, for different noise models. Our results show that our proposed decoding method achieves $\sim10 \times$ and $\sim2 \times$ higher values of pseudo-threshold and threshold respectively, than for MWPM. We show that usage of more sophisticated ML models with higher training/testing time, do not provide significant improvement in the decoder performance. Finally, data generation for training the ML decoder requires significant overhead hence lower volume of training data is desirable. We have shown that our decoder maintains a good performance with the train-test-ratio as low as $40:60$.
\end{abstract}



\section{ Introduction}

Quantum computers are expected to provide faster and often more accurate solutions to some of the problems of interest such as factorization \cite{Shor:1997:PAP:264393.264406}, database searching \cite{Grover:1996:FQM:237814.237866}, Hamiltonian simulation \cite{childs2012hamiltonian}, finding the lowest energy configuration of molecular systems \cite{mcclean2016theory}.  At any instant, while a bit in a classical computer can either be 0 or 1, a quantum bit known as a qubit is a unit vector in a 2-D complex vector space known as Hilbert Space. A qubit can be in state $\ket{0} = (1 \quad 0)^T$ or state $\ket{1} = (0 \quad 1)^T$ as well as in any linear superposition of the two basis states $\ket{0}, \ket{1}$. In general, a quantum state can be defined as $\ket{\psi} = \alpha\ket{0} + \beta\ket{1}$, where $\alpha, \beta \in \mathbb{C}$, $|\alpha|^2 + |\beta|^2 = 1$. Further, a quantum operation on qubits is reversible and can be represented by a unitary matrix.

Quantum states are however very prone to errors. Being vectors in Hilbert Space, even the slightest unwanted rotation occurring due to interaction with the environment, introduces error in the quantum system. It was shown by Shor \cite{PhysRevA.52.R2493} that any quantum error, also an operator, can be expressed as a linear combination of the Pauli matrices ($I = \begin{pmatrix}
1 & 0\\
0 & 1
\end{pmatrix}$, $X = \begin{pmatrix}
0 & 1\\
1 & 0
\end{pmatrix}$, $Z = \begin{pmatrix}
1 & 0\\
0 & -1
\end{pmatrix}$, and $Y = i \cdot Z \cdot X$). Therefore, if a quantum error correcting code (QECC) can correct the Pauli errors, then it can also correct any unitary error on the system. Researches have been carried out to design efficient QECCs; Shor first provided the 9-qubit code \cite{PhysRevA.52.R2493}, Steane improved it to 7-qubit code \cite{PhysRevLett.77.793} and finally Laflamme showed that 5-qubit code is optimum in the number of qubits \cite{PhysRevLett.77.198}.

In the circuit realization of the above-mentioned QECCs, there are multiple operations involving qubits which are not adjacent to each other. Operation on two far-apart qubits is costly since it requires multiple swap operations making the computation both slow and error-prone. Surface code was introduced to solve this drawback, called the Nearest Neighbour (NN) problem, where the qubits are placed in a 2D grid-like structure \cite{bravyi1998quantum, dennis2002topological,fowler2009high,wang2011surface,wootton2012high}, and the operations for error correction are only performed between adjacent qubits.

Quantum error correction includes the concept of encoding and decoding where the former creates a more secure logical qubit using multiple physical qubits, and the latter identifies the type and location of error. A {\em logical qubit} can be represented as $|q_L> = ENC(\displaystyle \otimes_{i=1}^{n} q_i)$ where ENC is the encoding technique and $q_i$ are the physical qubits.
A mathematical notion of quantum error correction using the stabilizer formulation has been provided in \cite{gottesman1997stabilizer}. Stabilizers are a set of operators such that $|q_L>$ is a $+1$ eigen state of the stabilizers (vide Section II for formal definition). A QECC can correct those errors occurring on $|q_i> $ for which $|q_L> $ becomes a  $-1$ eigen state of the stabilizers.  

The weight of an error operator $e$ is the number of non-identity Pauli operators present in its linear combination. The distance $d$ of a QECC is the minimum weight of an error that commutes with all the stabilizers. Therefore, if an error of weight $d$ or more occurs on the physical qubits that remain undetected by the stabilizers, it corrupts the logical qubit as well. Such an error is termed as {\em logical error}. Two parameters are used to evaluate the performance of a decoder \cite{fowler2012towards}: (i) the {\em pseudo-threshold}, which is the probability of physical error below which error-correction leads to a lower logical error probability, and (ii) the {\em threshold}, which is the probability of physical error beyond which increasing the distance of the code leads to higher logical error probability.

With increasing code distance,  the pseudo-threshold for a particular decoder also increases, which supports the intuition that using larger distance gives better protection from noise. On the other hand, the threshold does not change with respect to the distance because a  decoder for a particular surface code yields a fixed threshold.
The higher are the values of these parameters, the better is the performance of the decoder. Of these two parameters, the pseudo threshold is lower than the threshold for a decoder. The reason is that error correction is effective below the pseudo-threshold point, and coding theory asserts \cite{hill1986first} that in this region, increasing the distance of the code leads to higher suppression of logical errors. Therefore, if the threshold point is below the pseudo-threshold point it violates coding theory.
Hence, a decoder needs to have a higher pseudo-threshold than a higher threshold, since, beyond this error probability, QECC no longer provides any improvement in suppression of errors.
 
Let there be $n$ physical qubits in a logical qubit (eg. $n = d^2$ for surface code). A logical error can occur only when at least $d$ of the $n$ physical qubits are erroneous. Nevertheless, the presence of $d$ or more physical errors does not necessarily imply the presence of a logical error (see Section II for more details). If $p$ and $p_L$ are respectively the probability of physical and logical error, then
\begin{center}
    $p_L \leq \displaystyle \sum_{i=d}^n p^i $
\end{center}
Moreover, incorrect decoding itself can lead to logical errors. This can happen when the decoder fails to detect the actual physical errors and thus incorporates more errors during correction. Once again, not every incorrect decoding leads to a logical error (see Section II for more details). Therefore, if $p_d$ is the probability of failure of the decoder, then 
\begin{center}
    $p_L \leq \displaystyle \sum_{i=d}^n p^i + f(p_d)$
\end{center}
where $f(p_d)$ is a function of the probability of failure of the decoder. The function $ f(p_d)$ may vary with the decoder, hence the logical error probability may differ, resulting in different values of pseudo-threshold and threshold.

The performance of the decoder used is of utmost importance for any QECC. The performance of a decoder is measured in terms of both the accuracy and time requirement. In a fault-tolerant quantum computer, the qubits are encoded only once at the beginning of the computation, whereas they are decoded multiple times during the computation. Therefore, the decoding time is critical since a decoder taking longer time slows down the computation \cite{nielsen2002quantum, majumdar2016error}. The most popular decoding algorithm is Blossom Decoder \cite{edmonds_1965} which uses the Minimum Weight Perfect Matching (MWPM) algorithm, where the decoding time increases as $\mathcal{O}$(N$^4$) where $N$ is the number of qubits. Recently machine learning (ML) is being used for decoding purposes \cite{chamberland2018deep,varsamopoulos2017decoding,krastanov2017deep, sweke2018reinforcement} where, once the system is trained, the decoding technically runs as $\mathcal{O}(1)$. Moreover, MWPM works satisfactorily when the error probability of the system is low, as it always tries to find the minimum number of errors that can lead to the syndrome (discussed in detail in the Section II) at hand. But it does not consider the error probability of the system during decoding, which ML does. 

Multiple errors in a surface code can cause the same syndrome. An ML-based decoder may fail to distinguish between two such errors, and thus creating a logical error. In other words, the decoder may incorporate logical errors while correcting physical errors. Therefore, two stages of decoding, a low level followed by a high level, have been proposed in the literature for surface code. The former rectifies physical errors from syndromes and the latter checks for logical errors that might have been incorporated during the correction.

Most of the studies in surface code decoding consider only low-level decoders based on traditional decoding methods. In \cite{varsamopoulos2017decoding} the authors have considered ML-based decoders in conjunction with traditional decoders. In \cite{varsamopoulos2019comparing} the authors have used ML for both low and high-level decoders. However, their noise model is not well defined, since the authors did not mention whether they consider errors in a single or multiple steps in the error correction cycle of surface code (see Fig: \ref{fig:SCODE2}). Moreover, a question that has remained unanswered in previously studied ML-based decoders is whether using more sophisticated ML models can significantly enhance the performance of the decoder.

The main contributions of our work here can be summarized as follows:
\begin{itemize}
    
     \item
   We study the performance of ML based low and high level decoders, for distance 3, 5 and 7 surface codes, where error can occur in one or more of the eight steps in the surface code QECC cycle with equal probability. We show that our ML based decoder achieves $\sim 10 \times$ higher pseudo-threshold  and  $\sim 2 \times$ higher threshold as compared with MWPM based decoder.
    
    \item  We use Feed Forward Neural Network (FFNN) and Convolutional Neural Network (CNN) of varying sophistication (i.e., number of layers, nodes in each layer, etc.) and show that more sophisticated ML models cannot provide considerable improvement in the performance of the decoder, for distance 3, 5 and 7 surface codes. In fact, the time required for training the sophisticated ML models outweighs the minute improvement that is rarely or almost never observed.
    
     \item We empirically study the sample size required for training such a surface code decoder for different error probabilities and noise asymmetry because generation of a large training set imposes a significant overhead on the quantum device. An ML-decoder which requires a small number of training data, without hampering its performance, is more effective. We show that our decoder achieves its optimal performance for training size as low as $40\%$ of the entire dataset. This makes our ML-decoder a probable candidate for usage in near-future error-corrected quantum devices. We have also shown that the standard deviation of the performance of the ML-decoder increases with increasing physical error probability ($p_{phys}$). This supports the basic intuition that the steadiness of the performance of the ML decoder decreases with increasing $p_{phys}$. 
    
      \item We show that our ML-based decoder outperforms the one based on MWPM for varying degrees of asymmetry in the noise model as well. Real-world quantum channels are usually asymmetric \cite{PhysRevA.75.032345}, i.e., $Z$ errors are much more probable than $X$ or $Y$ errors. The performance of earlier ML-based decoders did not explore asymmetric noise models to the best of our knowledge.
\end{itemize}

 The rest of the paper is organized as follows: 
 In Section 2 we have discussed about the stabilizer formulation of the surface code. In Section 3 we have reviewed how machine learning techniques which are used in existing literature can be implemented for decoding purpose. In Section 4 we have discussed the result with different error models and in Section 5 we conclude.
 
\section{Stabilizer Formulation of the surface code}

Gottesman \cite{gottesman1997stabilizer} proposed the stabilizer formulation for error correction. A set of mutually commuting operators $M_1, \hdots, M_r$, where each $M_i \in \{I,X,Z,Y\}^{\otimes n}$, are said to stabilize an $n$ qubit quantum state $\ket{\psi}$ if $M_i\ket{\psi} = \ket{\psi}$, $\forall$ $i$ \cite{gottesman1997stabilizer}. An error $e$ is said to correctable by the QECC, if $\exists$ stabilizers $M_e \subseteq \{M_1, \hdots, M_r\}$, such that $M (e\ket{\psi}) = -e\ket{\psi}$, $\forall$ $M \in M_e$.

\begin{figure}[htb]
    \centering
    \includegraphics[height=6 cm, width= 12 cm] {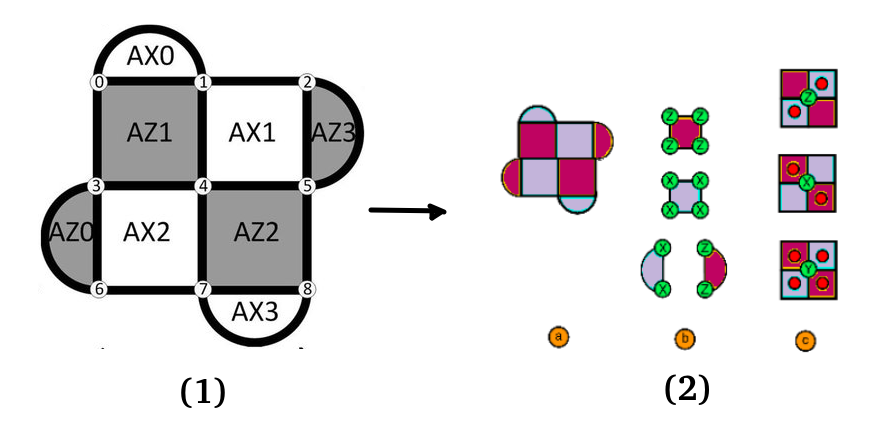}
    \caption{(1) Distance 3 surface code and (2) the syndromes. We consider a $d \times d$ lattice ($d$=3), with qubits on the vertices (referred to as physical qubits), and plaquette stabilizers (a). Pink (Purple) plaquettes indicate stabilizers which check the Z (X) parity of qubits on the vertices of the plaquette (b). Using red circles to indicate violated stabilizers we see some examples (c)\cite{varsamopoulos2017decoding} }
    \label{fig:SCODE}
\end{figure}

A QECC is called degenerate if $\exists$ errors $e_1 \neq e_2$ such that $e_1\ket{\psi} = e_2\ket{\psi}$. It is not possible to distinguish between such errors in a degenerate code. Surface code is a degenerate stabilizer code. Surface code is implemented on a two-dimensional array of physical qubits. The data qubits (in which the quantum information is stored) are placed on the vertices, and the faces are the stabilizers (refer Fig.~\ref{fig:SCODE}). The qubits associated with the stabilizers are also called measurement qubits. These are of two types --- $Measure$-$Z$~($M$-$Z$) and $Measure$-$X$ ~($M$-$X$). Each data qubit interacts with four measure qubits --- two $M$-$Z$ and two $M$-$X$, and each measure qubit in its turn interacts with four data qubits (Fig.~\ref{fig:SCODE}). An $M$-$Z$~($M$-$X$) qubit forces its neighboring data qubits $a$, $b$, $c$ and $d$ into an eigenstate of the operator product $Z_a Z_b Z_c Z_d$ ($X_a X_b X_c X_d$), where $Z_i$ ($X_i$) implies $Z$ ($X$) measurement on qubit $i$. Pauli-$X$ and Pauli-$Z$ errors are detected by the $Z$- and $X$- stabilizers respectively (Fig.~\ref{fig:SCODE}). An $X$~($Z$) logical operator is any continuous string of $X$~($Z$) errors that connect the top (left) and bottom (right) boundaries of the 2D array. The number of measure qubits, and hence the number of stabilizers, is one less than the number of data qubits when encoding a single logical qubit of information. An error-correcting code can correct up to $t$ errors if its distance $d \geq 2t+1$. A distance 3 surface code consists of 9 data qubits and 8 measure qubits (Fig.~\ref{fig:SCODE}). Thus a total of 17 qubits encode a single logical qubit, and hence the distance 3 surface code is also called SC17. 

\begin{figure}[htb]
    \centering
    \includegraphics[height=3 cm, width = 9 cm] {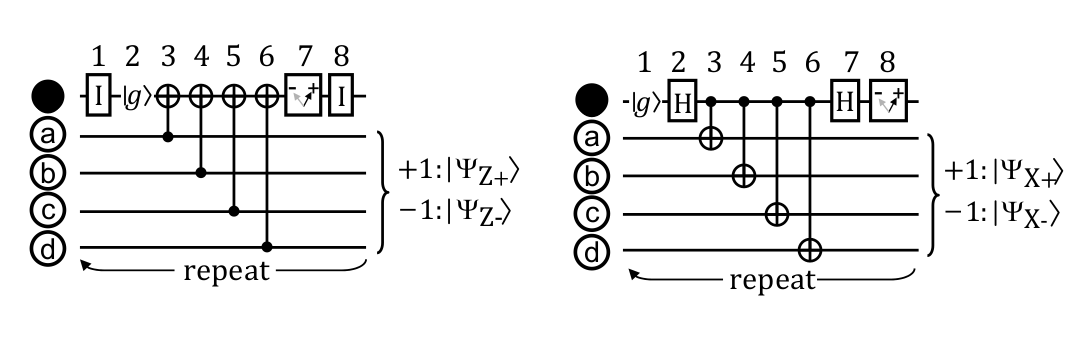}
    \caption{Quantum circuit for a single cycle of surface code for an M-Z and M-X qubit \cite{fowler2012towards}}
    \label{fig:SCODE2}
\end{figure}

\begin{figure}[htb]
    \centering
    \includegraphics[height=3 cm, width= 6 cm] {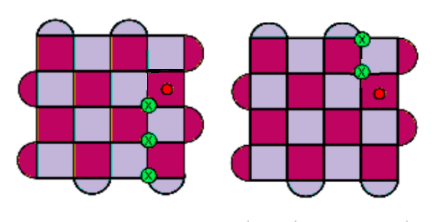}
    \caption{Example of surface code for d=5, where two errors produce the same syndrome \cite{varsamopoulos2017decoding}}
    \label{fig:dec}
\end{figure}

The circuit representations of the decoding corresponding to a single $M$-$Z$ and $M$-$X$ qubit are shown in Fig.~\ref{fig:SCODE2}. Since the same measure-qubit is shared by multiple data qubits, different errors can lead to the same syndrome in surface code. Hence the mapping from syndrome to error is not one-to-one, as illustrated in Fig.~\ref{fig:dec}. This often leads to poor decoding performance by decoders. In fact, if a decoder misjudges an error $e_1$ for some other error $e_2$, it can so happen that $e_1 \oplus e_2$ leads to a logical error. Therefore, not only the presence of physical errors, but also incorrect decoding can lead to uncorrectable logical errors as well. The goal of designing a decoder, thus, is to reduce the probability of logical error for some physical error probability.

\section{Machine Learning based decoding in surface code}

\subsection{Advantages of using Machine Learning based decoder}

Classical algorithms for decoding, such as Minimum Weight Perfect Matching (MWPM), are shown to perform poorly in certain scenarios. For example, MWPM tries to find the minimum number of errors that can recreate the obtained error syndrome without taking into consideration the probability of error. Furthermore, the time complexity of MWPM grows as $\mathcal{O}(N^4)$ where $N$ is the number of qubits. Lookup Tables have been used for decoding as well \cite{varsamopoulos2018designing}. While their performance is sometimes better than MWPM, its complexity scales as $\mathcal{O}(4^N)$ which becomes infeasible even for moderate values of $N$. To overcome such drawbacks, ML techniques have been applied to learn the probability of error in the system and propose the best possible correction accordingly with comparatively lower time complexity \cite{baireuther2018machine,varsamopoulos2017decoding,krastanov2017deep}. Supervised learning techniques, such as Feed-forward neural network (FFNN), Recurrent Neural Network (RNN) show that these are capable of outperforming the traditional decoding techniques.

\begin{figure}
    \centering
    \includegraphics[height=7.5 cm, width= 16.5 cm] {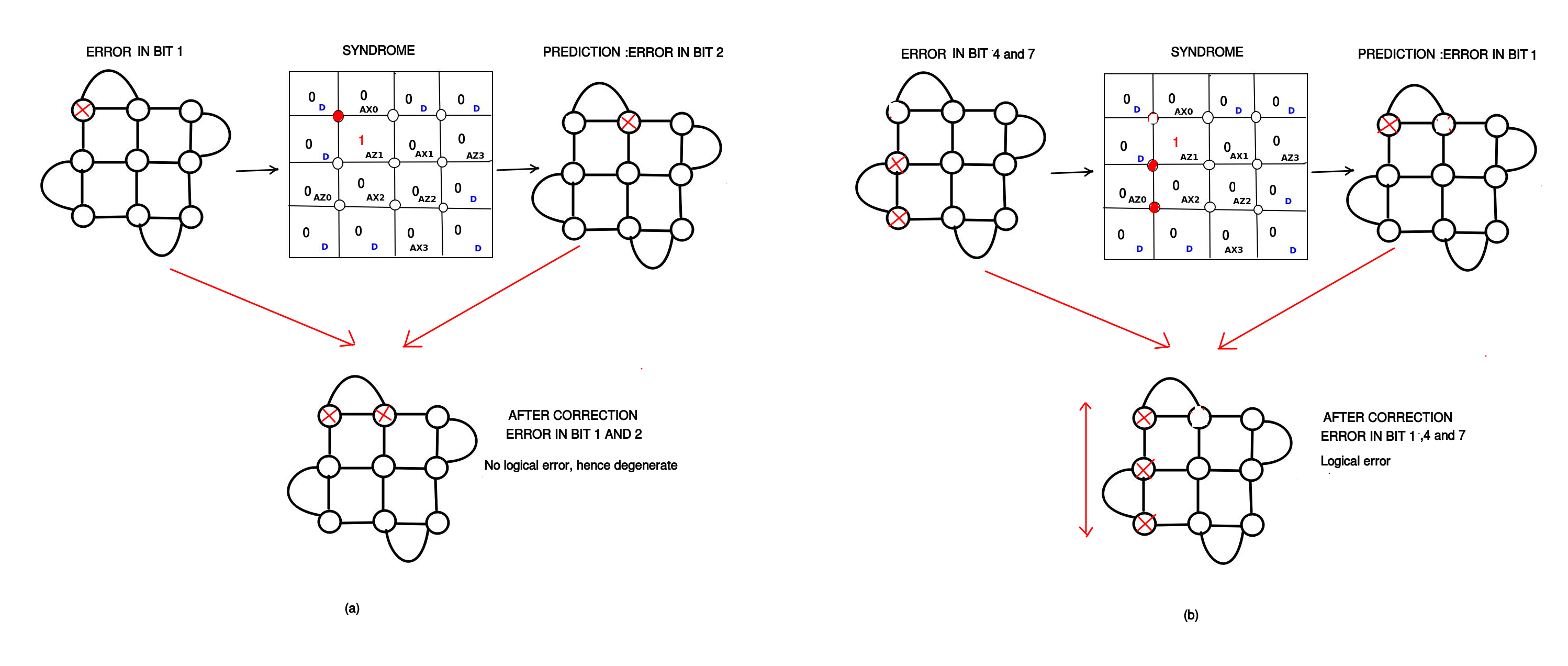}
    \caption{(a) No logical error and (b) Logical error due to mis-classification in low level decoding}
    \label{fig:logicalerror}
\end{figure}

As discussed earlier, surface code is degenerate, i.e., $\exists$ errors $e_1 \neq e_2$ such that $e_1\ket{\psi} = e_2\ket{\psi}$, where $\ket{\psi}$ is the codeword. This leads to any decoder failing to distinguish between some errors $e_1$ and $e_2$. Nevertheless, that does not always lead to a logical error. For example, bit-flip error in bit 1 and bit 2 are indistinguishable. But error in decoding these two will not lead to a logical error (Refer Fig.~\ref{fig:logicalerror} (a)). On the other hand, it is possible that $e_1 \oplus e_2$ leads to a logical error, i.e., the decoder may itself incorporate logical errors while correcting physical errors. For example, bit-flip error on qubit 4 is indistinguishable from those on qubits 1 and 7 together. But failure to distinguish between these two bit-flip errors leads to logical error (refer Fig.~\ref{fig:logicalerror} (b)). In general, \textit{usually} the decoder incorporates logical errors when it fails to distinguish between $\lfloor \frac{d-1}{2} \rfloor$ and $\lceil \frac{d+1}{2} \rceil$ errors. Broadly speaking, ML can learn the probability of error and predict which of those two are more likely. This makes ML-decoder outperform other traditional decoders.

Since a decoder itself can incorporate logical errors, two stages of decoders, namely low level followed by high level decoder, have been proposed in the literature for surface code \cite{varsamopoulos2019comparing}.

\begin{enumerate}
    \item  Low level decoders search for exact position of errors at the physical level.
    \item  High level decoders attempt to correct any logical error incorporated by the correction mechanism of low level decoders.
    
\end{enumerate}

\subsubsection{Design methodology of the ML based decoder}
Artificial neural networks (ANN) are made to emulate the way human brains learn, and are one of the most widely used tools in ML. Neural networks consist of one input layer, one output layer, and one or more hidden layers consisting of units that transform the input into intermediate values from which the output layer can find patterns that are too complex for a human programmer to teach the machine. The time complexity of training a neural network with $N$ inputs, $M$ outputs and $L$ hidden layers is $\mathcal{O}(N\cdot M \cdot L)$. Once the network is trained, testing is essentially done in $\mathcal{O}(1)$. In this paper we are using neural networks as both low-level and high-level decoder for distance 3, 5, and 7 surface code.

To apply ML techniques to surface code decoding, we first map the decoding problem to a common problem in machine learning, i.e. classification. Given a set of data points, a classification algorithm predicts the class label of each data point. In the following portion, we describe in detail the generation and training of ML as a decoder for surface code.

\subsubsection{Surface Code to a square lattice}
\begin{figure}[htb]
    \centering
    \includegraphics[height=7.5 cm, width= 12 cm] {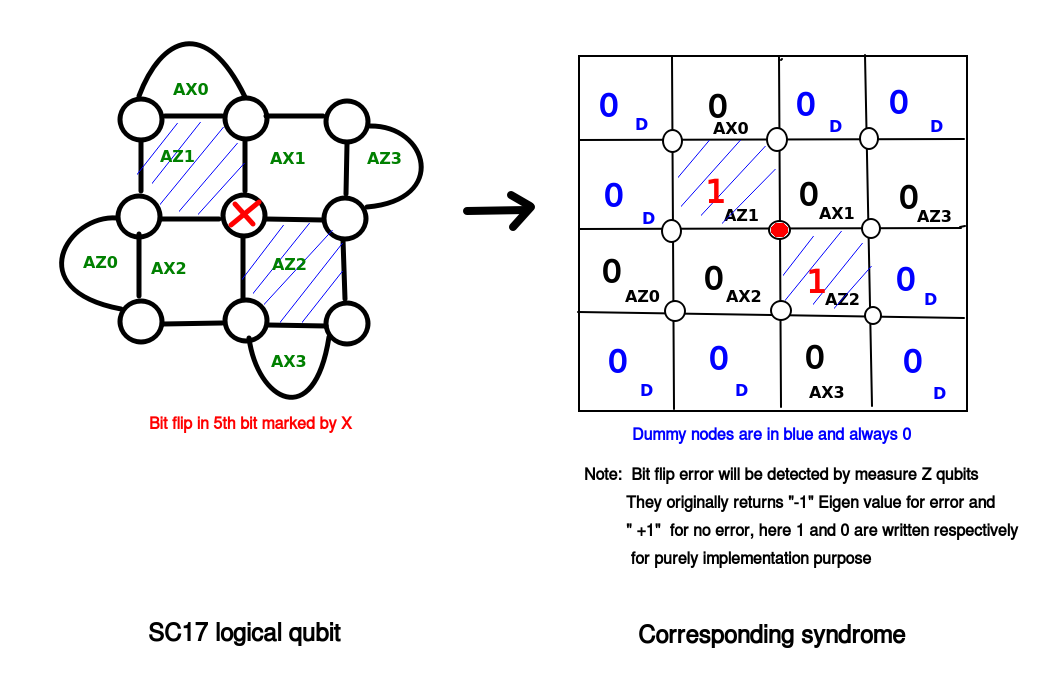}
    \caption{SC17 to syndrome generation}
    \label{fig:qubittosyndrome}
    \vspace{-0.5cm}
\end{figure}

For ease of implementation, we map the surface code to a square lattice (refer Fig.~\ref{fig:qubittosyndrome}). This is done by padding a few dummy nodes (labelled as $0_D$ in the figure). A distance $d$ surface code is converted into a $(d+1) \times (d+1)$ square lattice which has $d^2 - 1$ stabilizers, when encoding a single logical qubit. Therefore, $2(d+1)$ dummy nodes are required for this square lattice. The dummy nodes are basically don't care nodes, and their value is always 0 irrespective of the error in the surface code. The syndrome changes the values of the stabilizers only. 

\subsubsection{Error injection and syndrome extraction}

  \begin{figure}[p]
    \centering
    \includegraphics[height=22 cm, width= 10cm] {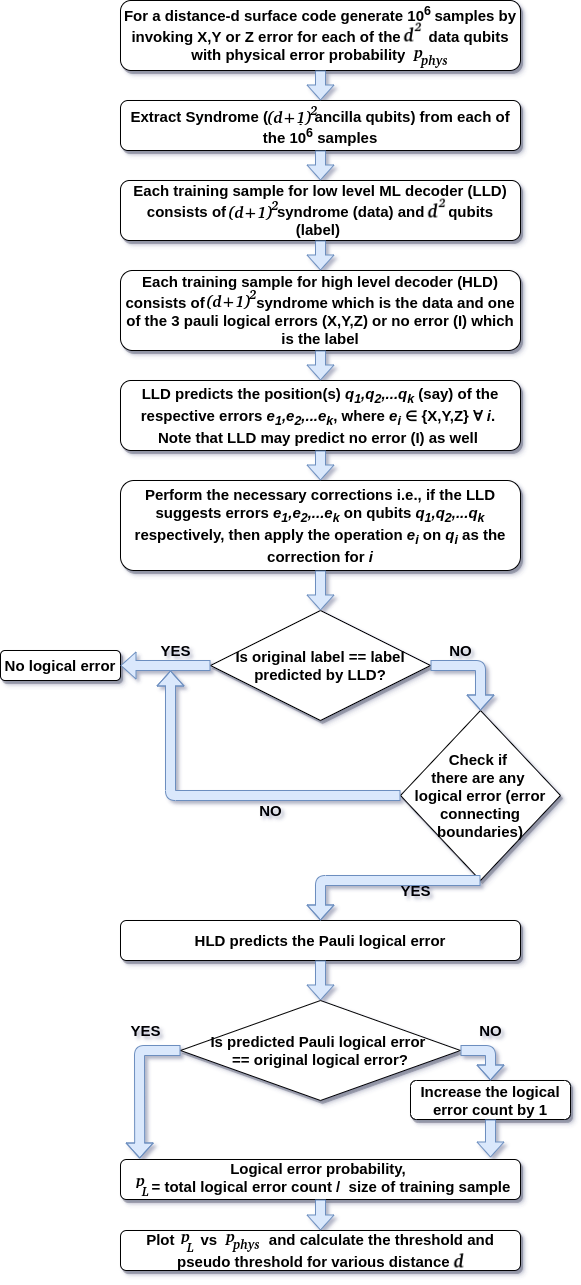}
    \caption{Outline of the ML based decoding}
    \label{fig:flow}
\end{figure}

Once the distance $d$ surface code is transformed to a $(d+1) \times (d+1)$ square lattice, the next step is to extract the syndrome for errors. First, we create a training dataset, where in each data we randomly generate errors on each physical qubit. If $p_{phys}$ is the probability of error on a physical qubit, the total probability of error after the 8 steps of surface code cycle (Fig.~\ref{fig:SCODE2}) is $1 - (1-p_{phys})^8$. We have trained the networks with $p_{phys}$ ranging from 0.0001 to 0.25. For generating the training data we have considered bit flip errors, symmetric and asymmetric depolarizing noise models. We have not separately considered phase flip errors since they are similar to bit flips and have a rotational symmetry (i.e., the logical errors of bit flip and phase flip model are equivalent up to a rotation by $\frac{\pi}{2}$).

From the training data (which may or may not contain errors), we generate the syndrome (measured by ancilla qubits of the surface code) (Fig.~\ref{fig:qubittosyndrome}). The syndrome, in our implementation, contains both the ancilla and the dummy nodes. However, the dummy nodes are always 0, whereas the values of the ancilla changes with different errors. Henceforth, in terms of implementation only, syndrome for a distance $d$ surface code will imply $(d+1)^2$ values including ancilla and dummy nodes. The final training data contains the syndrome, and its corresponding label is the true set of errors that have occurred in the system. Note that this method can lead to multiple labels having the same syndrome. This agrees with the fact that surface code does not have one-to-one mapping from error to syndrome.

Ideally, the dataset to achieve the best decoding performance should include all possible error syndromes. But as the code distance increases, the state space also increases exponentially. Therefore, we can at most include only a small percentage of the entire input dataset. The dataset size that we have used is 100000 from which 70000 is used for training and the rest for testing purpose.

\subsubsection{Training the ML model}

For the low level decoder, we train a neural network where the input layer is the syndrome and the output layer denotes the types of errors along with the physical data qubit where each error has occurred. For a distance $d$ surface code, the number of input nodes is $(d + 1)^2$ containing $d^2 - 1$ measure qubits and $2(d+1)$ dummy nodes. For example, if we consider a distance-3 surface code (SC17), it has 8 ancilla qubits and 8 dummy nodes. Therefore, in the input layer, there are 16 nodes (Fig.~\ref{fig:qubittosyndrome}).
In the output layer, there are 2 nodes for each data qubit to differentiate among $I$, $X$, $Y$ and $Z$ errors. The size of the hidden layer can be adjusted by trial-and-error. 

We have used two types of neural networks, (i) Feed Forward Neural Network (FFNN) and (ii) Convolutional Neural Network (CNN). For FFNN, we have taken 2 hidden layers having 32 and 16 nodes respectively. For the cost function we have used the mean squared error rate, and as the activation function we have used Rectified Linear Unit (ReLU). 

For CNN, the first layer is a 64 dimension convolution layer where input is a $4 \times 4$ matrix and the kernel size is also $4 \times 4$. Then we flatten it and add two fully connected layers of dimension 64 and 32. After that we add the fully connected output layer of dimension 9. For the first 3 layers (convolution, dense, dense) we have used ReLu as the activation function and for the output layer we have used sigmoid activation function since it will be a multi-label classification problem. These values were adjusted after multiple trial-and-errors. We later show in the result section that building a more complex neural network cannot provide any significant increase in the performance of the decoder, but requires significantly more decoding time. Therefore, we stick to these parameters.

The high-level decoder simply tries to predict any logical error that has been incorporated by the low level decoder. Therefore, its input remains the same as the low-level decoder (i.e., the syndrome) whereas it has 4 nodes in the output, each corresponding to a logical Pauli operator.
 
First, the network is trained for low-level decoder. After the low-level decoding is done, the predicted corrections are applied, and rechecked by  using the high-level decoder whether any logical error has been inserted by the low-level decoder. The entire workflow is given in Fig.~\ref{fig:flow}.

\section{Results}
First, we focus on the decoding performance of an ML-based
low-level and the high-level decoder for surface codes of distances 3, 5, and 7 for both symmetric and asymmetric depolarizing noise models with varying degrees of asymmetry. Our model outperforms the performance of the existing decoders for symmetric noise model. We also show that although the performance of ML is slightly poorer for asymmetric noise models than that for the symmetric one, it still outperforms MWPM. Furthermore, we provide an empirical study to estimate the minimum train-test-ratio needed for optimal accuracy to obtain a better estimate of the minimum number of training data required to obtain the best (or near best) decoding results with ML decoder.

In the following subsections, we first introduce the noise model that we have considered, followed by the parameters of our ML decoder. Finally, we show the results of our decoder and compare its performance with the traditional MWPM decoder.

\subsection{Noise models}

Given a quantum state $\rho$ in its density matrix formulation \cite{nielsen2002quantum}, the evolution of the state in a depolarization noise model is given as

$$\rho \rightarrow (1-p_x-p_y-p_z)\rho + p_x X \rho X^{\dagger} + p_y Y \rho Y^{\dagger} + p_Z Z \rho Z^{\dagger}$$

where $p_x, p_y, p_z$ represent the probability of occurrence of unwanted Pauli $X$, $Y$, and $Z$ error. In symmetric depolarization noise model, $p_x = p_y = p_z$. Moreover, quantum channels are often asymmetric or biased, i.e., the probability of occurrence of $Z$ error is much higher than that of $X$ or $Y$ error.
Furthermore, each error correction cycle in surface code requires eight steps. We have considered that an error can occur on one or more of the $d^2$ data qubits in each of the eight steps, where $d$ is the distance of the surface code. Therefore, if $p_x + p_y + p_z = p$, then the overall probability of error for each error correction cycle is $1 - (1-p)^8$. We assume noise-free measure qubits and ideal measurements.

\subsection{Machine Learning Parameters}

For our study, we have trained the ML model with batches of data, not the entire data set at once. This is often beneficial in terms of training time as well as memory capacity. We have used batch size = 10000, epochs = 1000, learning rate = 0.01 (with Stochastic Gradient Descent), and we have reported the average performance of each batch over 5 instances. This method is repeated for each value of the $p_{phys}$ considered in this study.

\begin{figure}
     \centering
     \begin{subfigure}[b]{\textwidth}
     \centering
         \includegraphics[scale=0.27] {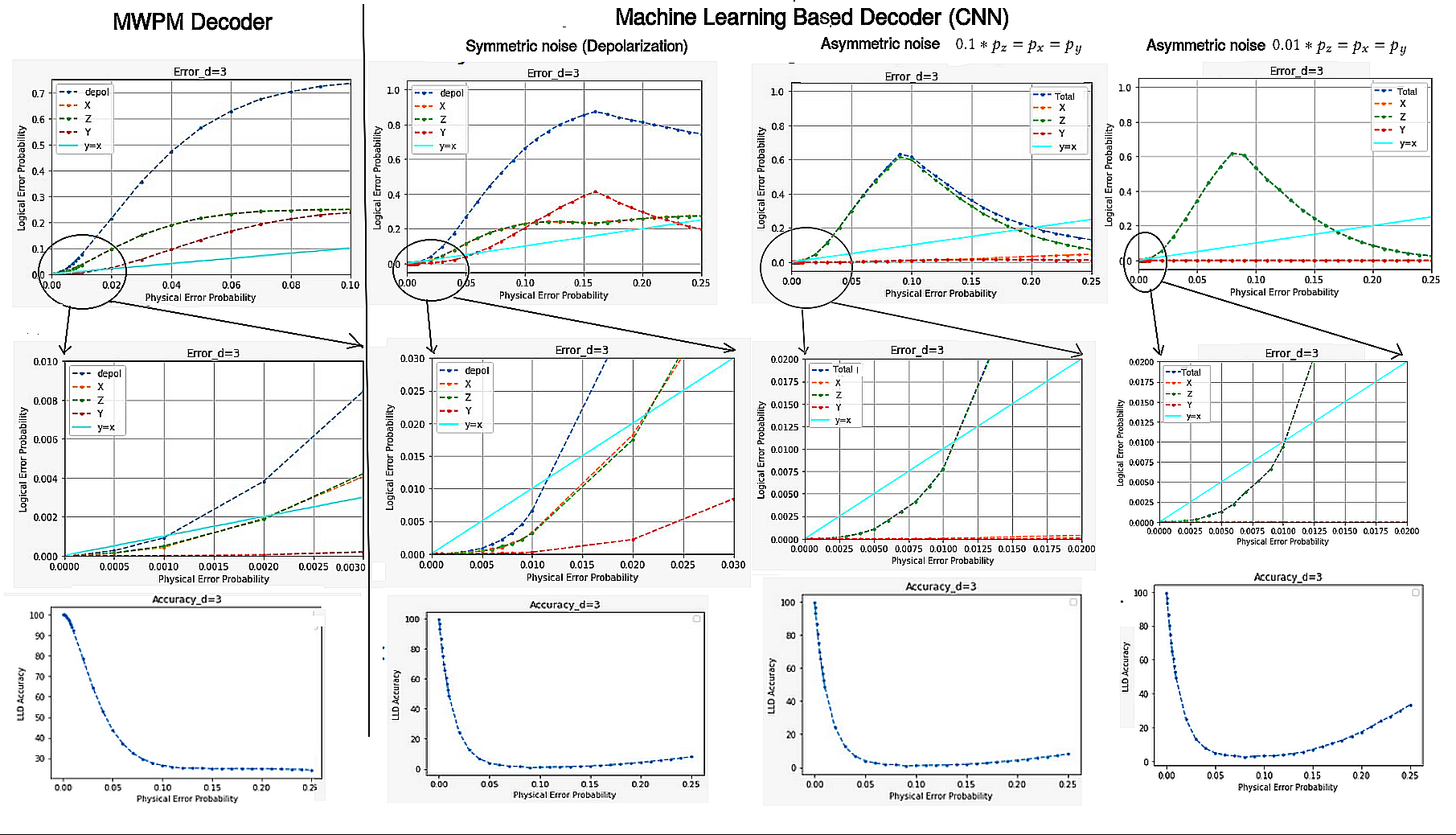}
         \caption{Pseudo-threshold and accuracy --- MWPM vs ML-based decoder for distance 3 surface code }
         \label{result1MWPMvsMLD3}
     \end{subfigure}
     
     \begin{subfigure}[b]{\textwidth}
     \centering
         \includegraphics[scale=0.27] {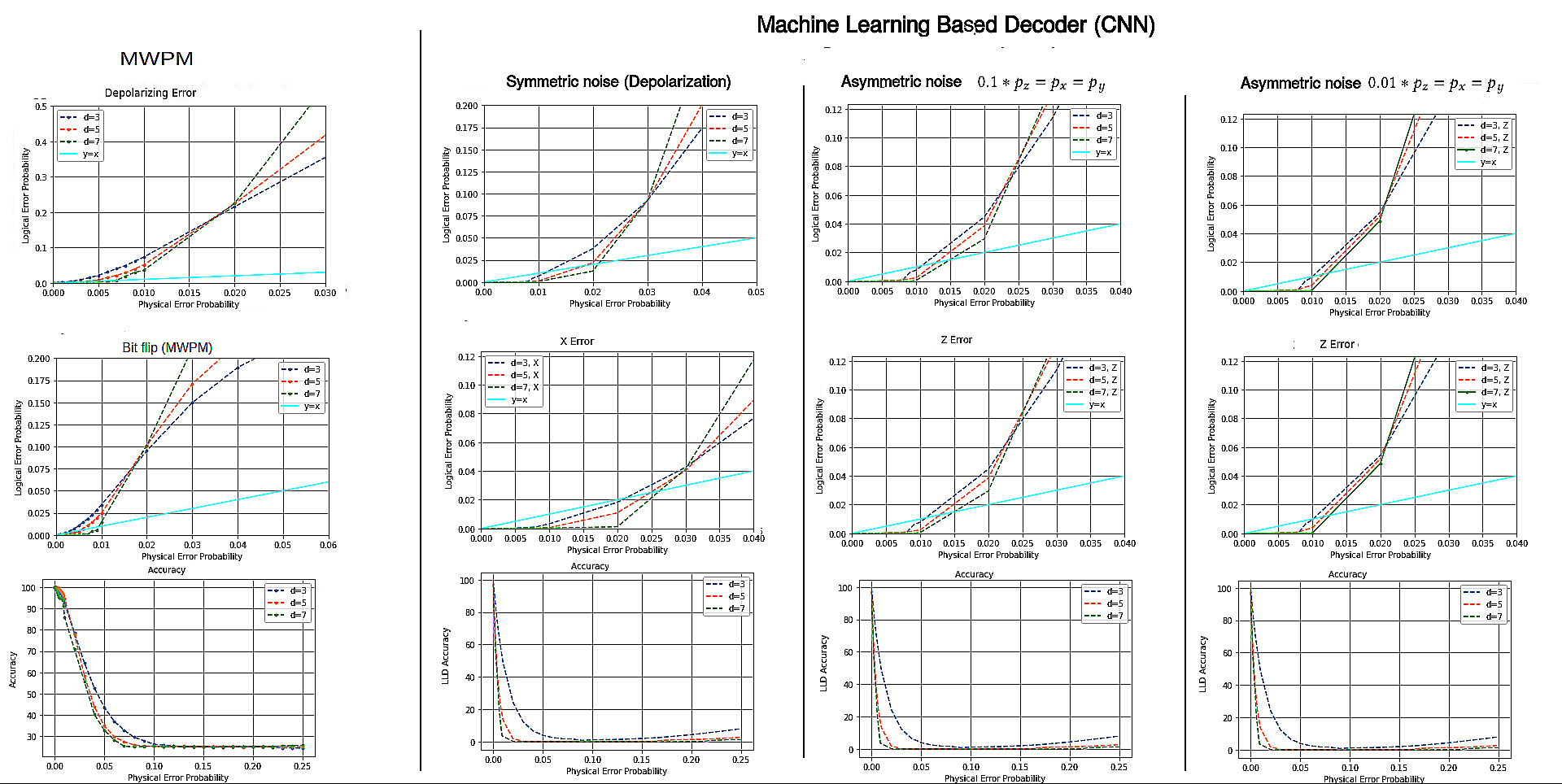}
         \caption{Threshold and accuracy --- MWPM and ML-based decoder for surface code with $d$ = 3, 5, 7}
         \label{result1MWPMvsMLD3and5}
     \end{subfigure}
     
    \caption{Comparison of pseudo-threshold, accuracy and threshold for MWPM vs. ML-based decoder }
    \label{fig:result1MWPMvsML}
\end{figure}

\subsubsection{Low and high-level decoder}
In Fig.~\ref{fig:result1MWPMvsML}, we show the increase in the logical error probability with physical error probability $p$, which is the probability of error per step in the surface code cycle. The results of MWPM and CNN-based low-level decoder for both symmetric and asymmetric noise models are shown. In Tables~\ref{table:pseudothreshold} and ~\ref{table:threshold}, we depict the performance of FFNN decoder as well. In Fig.~\ref{fig:result1MWPMvsML}, the blue, yellow, green, and red lines respectively are the decoder curves which show the probabilities of logical error for symmetric depolarization, bit flip ($X$), phase flip ($Z$), and $Y$ errors. The cyan straight line consists of the points where the probabilities of physical and logical error are equal.

The point where the decoder curves and the straight line intersects, defines the value of pseudo-threshold for the decoder. As expected, the pseudo threshold improves with increasing distance of the surface code. Nevertheless, the threshold value is the probability of physical error beyond which increasing the distance leads to poorer performance. Therefore, threshold is independent of the distance and is a property of the surface code and the noise model only. In Tables \ref{table:pseudothreshold} and \ref{table:threshold}, we show the pseudo-threshold and threshold of the low and high-level decoders for distance 3, 5 and 7 surface code in symmetric and asymmetric noise models respectively. Fig.~\ref{fig:result1MWPMvsML} (a) shows the pseudo-thresholds for MWPM and CNN decoder for a distance 3 surface code using low-level decoder (LLD) only. From Table \ref{table:pseudothreshold} we observe $\sim 10\times $  increase in the pseudo threshold for ML-decoders as compared to MWPM.

Fig.~\ref{fig:result1MWPMvsML} (b) shows the thresholds and decoder accuracy for MWPM and ML-decoders surface codes of distance 3, 5 and 7. Table \ref{table:threshold} depicts the threshold values for MWPM and ML-decoders.  From Table \ref{table:threshold} we observe $\sim 2 \times $  increase in the  threshold for ML-decoders as compared to MWPM.

\begin{table}
    \centering
    \caption{Pseudo-Threshold of the low and high level decoders for distance 3, 5 and 7 surface code}
    \scriptsize
    \begin{tabular}{|c c|c|c|c|c|c|c|c|c|c|}
    \hline
        \multicolumn{2}{|c|}{\multirow{3}{*}{\backslashbox{Decoder}{Noise model}}} & \multicolumn{3}{c}{Symmetric} & \multicolumn{6}{|c|}{Asymmetric}\\
        \cline{3-11}
         & & \multicolumn{3}{c|}{$p_z = p_x = p_y$} & \multicolumn{3}{c|}{$0.1p_z = p_x = p_y$} & \multicolumn{3}{c|}{$0.01p_z = p_x = p_y$}\\
         \cline{3-11}
         \cline{3-11}
         & & d= 3 & 5 & 7 & 3 & 5 & 7 & 3 & 5 & 7\\
        \hline
        \multirow{2}{*}{MWPM} & LLD & 0.0011 & 0.0038 & 0.0075 & 0.0012& 0.0041 & 0.0072 & 0.00098& 0.0038& 0.0067 \\
        \cline{3-11}
        & HLD &- &- &- &- &- &- &- &- & -\\
        \hline
        \multirow{2}{*}{Our FFNN} & LLD & 0.012 & 0.0205& 0.0219 & 0.0109& 0.0121& 0.0152& 0.0120& 0.0122 & 0.0131 \\
        \cline{3-11}
        & HLD & 0.0143& 0.0234& 0.0241& 0.0124& 0.0164 & 0.0189& 0.0123& 0.0165 & 0.0189 \\
        \hline
        \multirow{2}{*}{Our CNN} & LLD & 0.0121 & 0.0211 &0.0228 & 0.0112 & 0.0125 & 0.0151& 0.0111&0.0121 &0.0132 \\
        \cline{3-11}
        & HLD &0.0152 &  0.0241 & 0.0247  &  0.0134 & 0.0161 &0.0192  &0.0121 & 0.0162 &  0.0195 \\
        \hline
    \end{tabular}
    \label{table:pseudothreshold}
\end{table}
\normalsize

\begin{table}
\centering
\caption{Comparison of Threshold  of the low and high level decoders}
\begin{tabular}{ |c|c|c|c| }
 \hline
Threshold (LLD)&Threshold (HLD)&Decoder Model&Error model\\
 \hline
0.0181 & N/A &MWPM & Symmetric \\
\hline
0.0302 & 0.035&  & Symmetric\\
0.0218 & 0.025& & Asymmetric $0.1 * p_z = p_x =  p_y$\\
0.0221 & 0.0279& Our FFNN & Asymmetric $0.07 * p_z = p_x =  p_y$\\
0.0216 &0.0257&  & Asymmetric $0.04 * p_z = p_x =  p_y$\\
0.0213 & 0.0251&  & Asymmetric $0.01 * p_z = p_x =  p_y$\\

\hline
0.0311 & 0.034& & Symmetric\\
0.0225 & 0.026&  & Asymmetric $0.1 * p_z = p_x =  p_y$\\
0.0229 & 0.0281& Our CNN & Asymmetric $0.07 * p_z = p_x =  p_y$\\
0.0223 & 0.0258&  & Asymmetric $0.04 * p_z = p_x =  p_y$\\
0.0212 &0.0252&  & Asymmetric $0.01 * p_z = p_x =  p_y$\\

 \hline
\end{tabular}

\label{table:threshold}
\end{table}

In Fig.~\ref{fig:result1MWPMvsML} (a), we observe that at very low error probability the accuracy remains good, then it falls drastically. However, for ML decoders, it again increases beyond a certain physical error probability (~0.15). On the other hand, the logical error also decreases in most of the cases for both symmetric and asymmetric ML decoders after more or less that same value of physical error probability. This is due to the bias in the back-end working principle of any machine learning model. When the error probability is low or high, the ML decoder effectively learns the probability and in most of the cases can avoid logical errors. But when the error probability is in the mid range, the ML model gets confused. For example, if in a training set, out of 12 events with same value of the features, 10 events are certainly in class A, and the rest in B, then the ML  definitely learns it with high accuracy. Similarly, if those same 10 events are in class B, accuracy will be high. But the ML is confused when 6 of them are in class A and 6 of them in class B. This is an interesting observation in the ML-decoder which is absent in MWPM-decoder.

\begin{figure}[!h]
     \centering
     \begin{subfigure}[b]{\textwidth}
     \centering
         \includegraphics[scale=0.4] {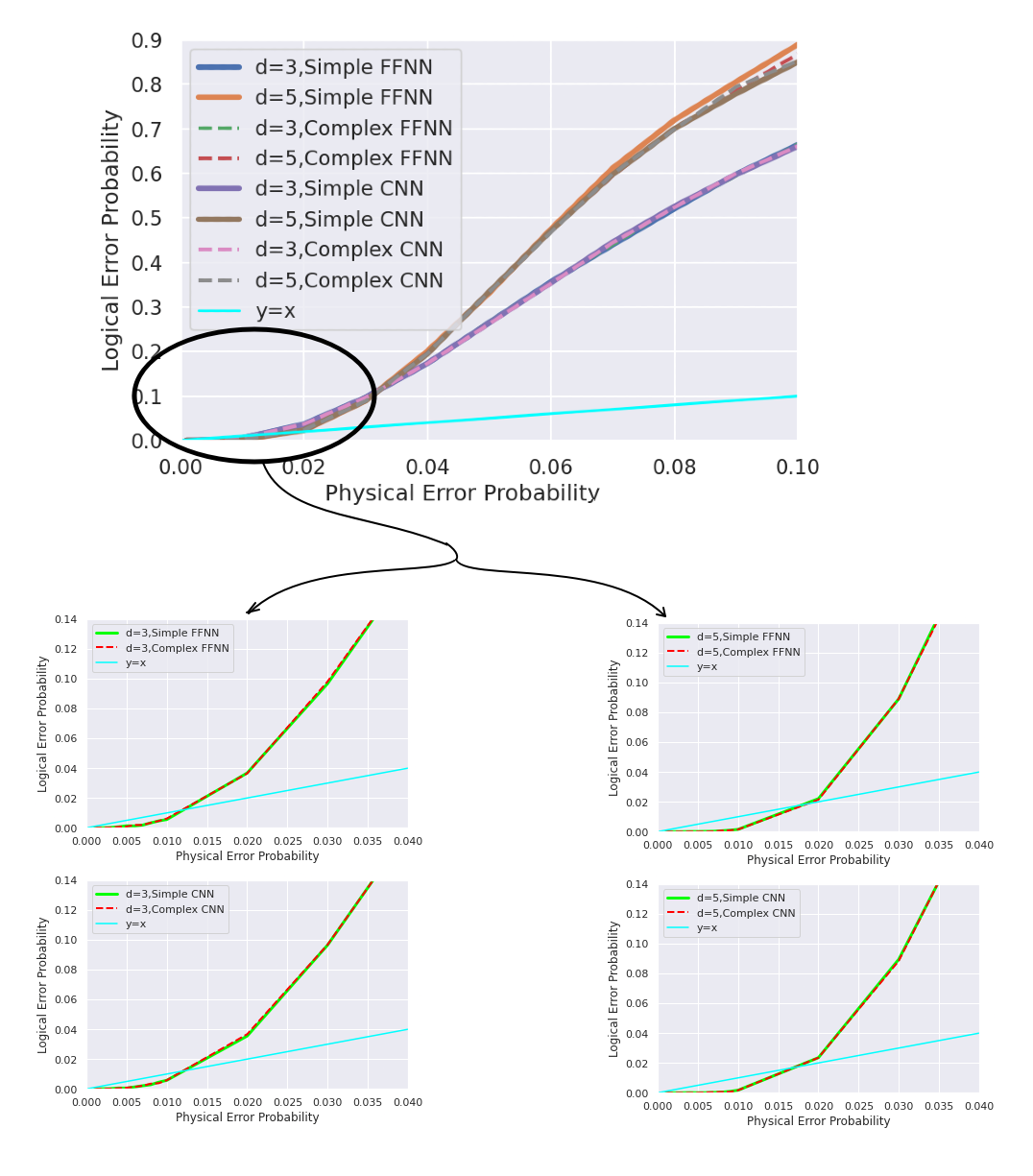}
         \caption{Logical vs physical error probability for various ML models in d= 3 and 5 surface code }
         \label{diffmodels}
     \end{subfigure}
     
     \begin{subfigure}[b]{\textwidth}
     \centering
         \includegraphics[scale=0.4] {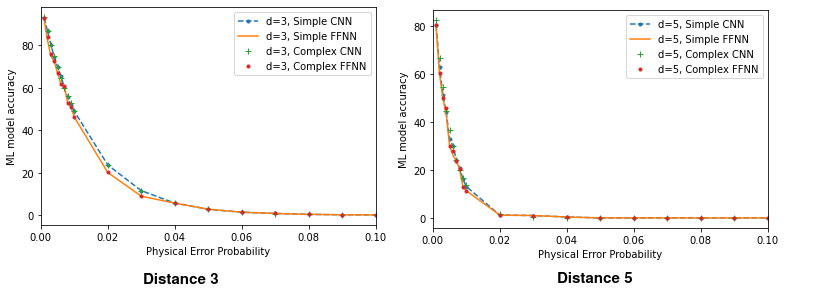}
         \caption{ML model accuracy vs physical error probability for various ML models in d = 3 and 5 surface code}
         \label{diffmodelsacc}
     \end{subfigure}

    \caption{Pseudo-threshold, threshold and accuracy of decoders with  different ML models}
    \label{fig:diffmodels_acc}
\end{figure}

\begin{table}[htb]
    \centering
    \caption{Comparison of training times for different ML models}
    \begin{tabular}{|c c|c|c|c|c|c|c|}
    \hline
         \multicolumn{2}{|c|}{\multirow{3}{*}{ML Model}} & \multicolumn{3}{c|}{d = 3} & \multicolumn{3}{c|}{d = 5}\\
         \cline{3-8}
         & & Parameter & Training & Prediction & Parameter & Training & Prediction\\
         & & space & time (sec) & time (sec) & space & time (sec) & time (sec)\\
         \hline
         \multirow{2}{*}{FFNN} & Simple & 2258 & 53.12 & $2.1 \times 10^{-5}$ & 5618 & 103.18 & $3.5 \times 10^{-5}$\\
         \cline{3-8}
         & Complex & 84754 & 324.9 & $3.55 \times 10^{-5}$ & 88114 & 394.99 & $3.72 \times 10^{-5}$\\
         \hline
         \multirow{2}{*}{CNN} & Simple &165650  &  785.27 & $5.27 \times 10^{-5}$ & 429874 & 1852.74 &  $7.4 \times 10^{-5}$  \\
         \cline{3-8}
         & Complex & 240246 & 1485.69 &  $6.02 \times 10^{-5}$ &  504370 & 4241.58  & $9.74 \times 10^{-5}$ \\
         \hline
    \end{tabular}
    \label{tab:my_label}
\end{table}

\subsection{More sophisticated ML models}

A natural question is whether the use of more sophisticated ML models (e.g, adding more hidden layers, increasing the number of nodes in each layer, etc.) can improve the performance of the decoder. We have addressed this issue as reported below.

\begin{figure}[!hb]
     \centering
     \begin{subfigure}[b]{\textwidth}
     \centering
         \includegraphics[scale=0.7] {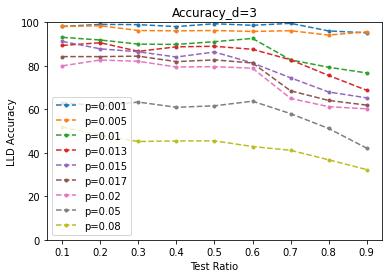}
         \caption {Average accuracy of our low level decoder}
         \label{traintestratio}
     \end{subfigure}
     
     \begin{subfigure}[b]{\textwidth}
     \centering
         \includegraphics[scale=0.6] {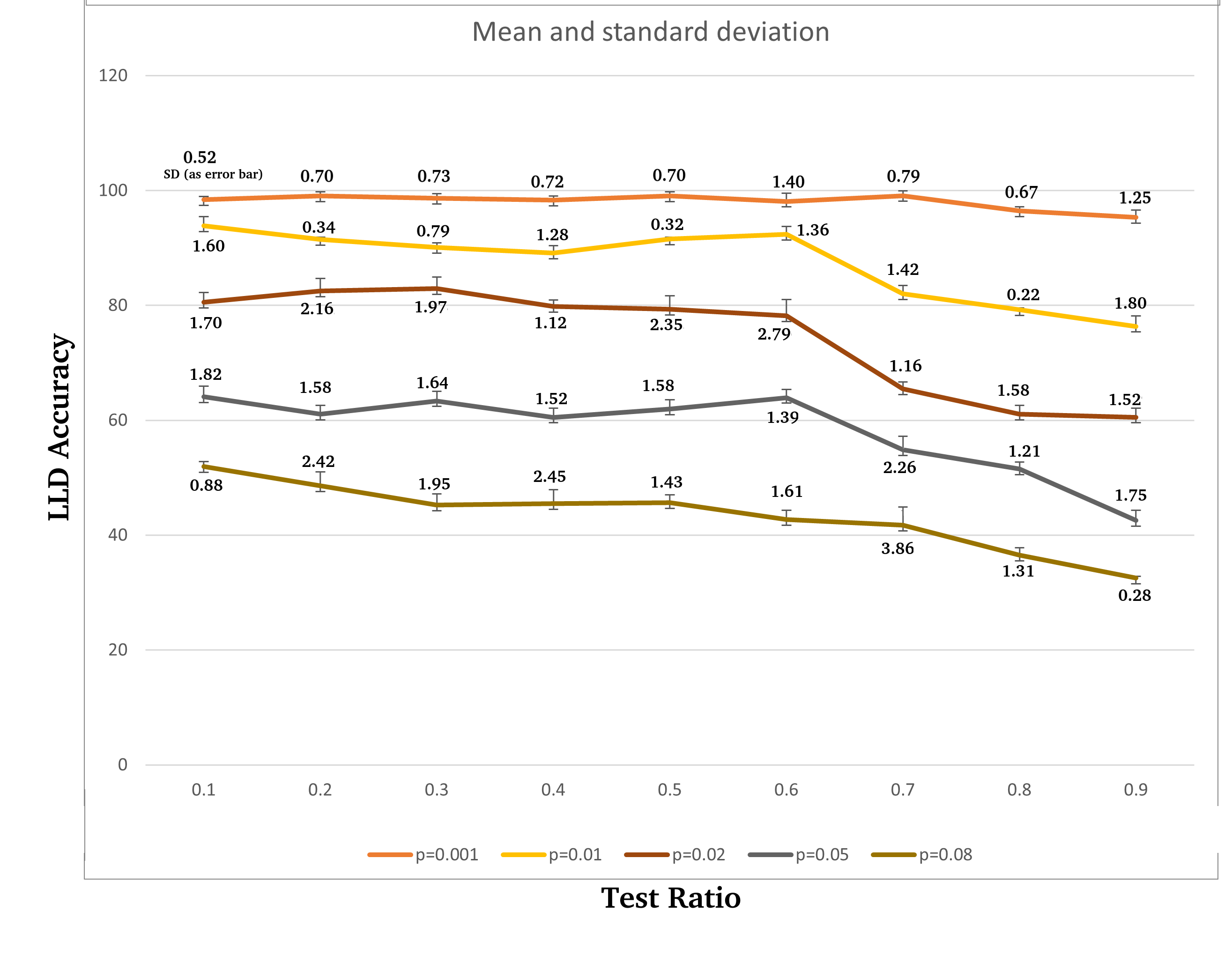}
         \caption{Average Accuracy with its standard deviation as error bar of our low level decoder}
         \label{traintestratio_sd}
     \end{subfigure}

    \caption {Average accuracy (along with its standard deviation) of our low level decoder vs Test Ratio for different values of $p_{phys}$ in distance 3 surface code}
    \label{fig:tt}
\end{figure}

In Table~\ref{tab:my_label}, Simple FFNN has 1 hidden layer (dense) whereas Complex FFNN has 5 hidden layers (dense) and Simple CNN has 1 convolution (64 dimensions) followed by 2 dense layers of dimension 256 and 64 respectively before the output layer whereas Complex CNN has 3 convolutions (64 dimensions) layers followed by 4 dense layers of dimension 512, 256, 128 and 64 respectively before the output layer. The more sophisticated models naturally require more time for training and prediction. But from Fig.~\ref{fig:diffmodels_acc}(a), we see that the decoder graphs are more or less overlapping for the simple and complex ML models. Therefore, it can be concluded that using more sophisticated model does not lead to a better performance for the decoder. This can be further verified by the accuracy plots in Fig.~\ref{fig:diffmodels_acc}(b). Since the more complex models are performing almost at par with the simpler models for $d=3$ and 5 and the complex models are significantly more time-consuming, we performed the experiments with $d = 7$ only for our Simple CNN and FFNN models.

\subsection{Empirical train-test-ratio for optimal accuracy}

In general, the higher the number of training samples, better is the accuracy of the ML model up to a certain threshold, beyond which increasing the number of training samples does not improve the performance of the model \cite{shalev2014understanding}. However, generation of training data is a humongous task in current quantum devices since it takes up a significant amount of device lifetime. Therefore, lower the size of the training sample required, higher is its usability. But naively reducing the size of the training set may lead to performance degradation. We explore this requirement by studying the minimum train-test-ratio required to obtain the optimal decoder performance.

In Fig.~\ref{fig:tt} we have varied the train-test ratio for the simple CNN decoder for a distance 3 surface code, starting from 90:10 and moving up to 10:90, lowering the training proportion by 10\% in each step, and have plotted the accuracy for increasing $p_{phys}$. We observe that even when we take only 60\% data as testing data, our model performs satisfactorily in case of different error probabilities. The performance degrades to some extent beyond this value. Therefore, we can conclude that this ML decoder can achieve its optimal performance even for quite a low size of training sample. For each test ratio and $p_{phys}$ the values in Fig.~\ref{fig:tt} were generated by averaging over 5 instances. 

Since this is a ML based method, and Fig.~\ref{fig:tt}(a) shows the mean value only, in Fig.~\ref{fig:tt}(b) we have also plotted the standard deviation (SD) with a few values of physical error probability for all the test-ratio as an \emph{error bar} plot. We observe that for $p=0.001$ the accuracy varies between 95.32 to 99.11 (min SD = 0.52, max SD = 1.40). For $p = 0.02$  the accuracy varies between 60.54 to 82.94 (min SD = 1.12, max SD = 2.79) and for $p = 0.08$  the accuracy varies between 32.51 to 51.92 (min SD = 0.28, max SD = 3.86). With increasing $p_{phys}$, the SD also increases. This supports intuition because as the $p_{phys}$ increases, the decoding performance decreases due to the capacity of the machine learning model to correctly classify the errors. Hence, the performance of ML (which depends on the errors in the dataset), varies more with higher value of physical error probability ($p_{phys}$). 

\section{Conclusion}

In this paper, we have proposed an ML-decoder to correct both symmetric and asymmetric depolarizing noise on surface codes. Our decoder has two levels --- in the low-level it tries to accurately predict the error on the qubits, followed by the high level that tries to detect any logical error that may have been introduced by the low-level decoder. Both these decoders have been implemented using neural network (FFNN and CNN) for surface code of distances 3, 5 and 7. Our proposed ML-decoder outperforms MWPM, and we observe $\sim 2\times $  increase in threshold and  $\sim 10\times $ increase in pseudo threshold. We further show that the decoder performance is equally good for asymmetric errors as well, which is more realistic in quantum devices.

We have used ML models with different levels of sophistication, (i.e. varying number of hidden layers and node-density of each layer). Our results show that the mere increase of complexity in ML model requires an increased amount of time for decoding but hardly yields any better performance. 

Finally, we show that this decoder can provide the optimum performance even with train-test-ratio as small as 40:60. At par with intuition, we find that the standard deviation of the performance of ML decoder increases with increasing physical error probability. The performance of our proposed decoder and its low requirement of training data makes it a viable candidate for decoding near-future surface codes in quantum computers.

In this work, we have assumed, noise-free measure qubits and ideal measurements. A future prospect of this research can be to consider noisy measure qubits and imperfect measurements.

\bibliographystyle{unsrt}
\bibliography{apssamp}

\end{document}